\documentclass[aps,prd,amssymb,twocolumn,nofootinbib]{revtex4}

\newcommand{\ie}{\mbox{i.\,e.\,\ }}
\newcommand{\iec}{\mbox{i.\,e.\,}}

\newcommand{\eg}{\mbox{e.\,g.\,\ }}
\newcommand{\egc}{\mbox{e.\,g.\,}}

\newcommand{\mc}[1]{\ensuremath{\mathcal{#1}}}

\newcommand{\pbp}[2]{\ensuremath{\frac{\partial #1}{\partial #2}}}

\newcommand{\be}{\begin{equation}}
\newcommand{\ee}{\end{equation}}

\begin{document}
\title{Fields as Bodies: a unified presentation of spacetime and internal gauge symmetry}
\author{David Wallace}
\email{david.wallace@balliol.ox.ac.uk}
\affiliation{Balliol College, University of Oxford}
\date{\today}
\begin{abstract}
Using the parametrised representation of field theory (in which the location in spacetime of a part of a field is itself represented by a map from the base manifold to Minkowski spacetime) I demonstrate that in both local and global cases, internal (Yang-Mills-type) and spacetime (Poincar\'{e}) symmetries can be treated precisely on a par, so that gravitational theories may be regarded as gauge theories in a completely standard sense.
\end{abstract}

\maketitle

\section{Introduction}

In its simplest formulation, a field theory is a function from a base space representing spacetime to some space of field values. A field is not so much an \emph{entity} as a way of talking about certain \emph{properties} of the points of spacetime. 

This is not simply a philosophical matter. Mathematically, field theories treat the spacetime and internal features of the theory --- and their associated symmetries ---  in sharply distinct ways. The spacetime structure of the theory is represented by structure on the base space (a metric, an affine connection and so forth); the internal structure is represented by structure on the field-value space.\footnote{Some accounts of general relativity, \eg \cite{mtw},\cite{MaudlinSpacetime} embrace this distinction, treating spacetime as the (dynamically active) container against which physics plays out, and gravity as the curvature of that background. Others (\eg \cite{rovelli,brownrelativity}) prefer to describe gravity as one more dynamical field, on a par with the fields of particle physics, and to downplay its distinctive aspects. I hope that the formalism of this paper provides a mathematical framework appropriate for the second conception.} 

This division persists when we move from global to local internal symmetries (the gauge principle) and from global to local theories of spacetime (from special to general relativity). Analogies are found between the spacetime and internal structures, but at a fundamental mathematical level they differ. An extensive literature on the degree to which general relativity is a gauge theory (and if so, a gauge theory of what?) points to the significance of this mathematical difference. In particular, so-called ``Poincar\'{e} gauge theory'' differs from Yang-Mills gauge theories in a number of respects and these differences are generally attributed to the different nature of spacetime and internal symmetries.\footnote{The first proposal for a gauge theory of gravity appears to have been \cite{utiyama} and the theory was developed further by \cite{Sciama1962} and \cite{kibble}; for technical details and further references see \cite{hehlblagojevicbook} and references therein.}

In this paper I demonstrate that the difference represents a certain choice in how a field theory is formalised, and not a difference of kind. It is in fact possible to formulate field theories --- local or global --- so that the spacetime features of the theory are treated in precisely the same way as the internal features. Conceptually, the key move is to stop thinking of a field as a collection of properties of spacetime, and to consider it instead as an extended body, whose parts have both spatiotemporal properties and internal-space properties. Mathematically, this move is \emph{parametrisation}: we remove the spacetime structure from the base space, and instead represent it as a further field --- the ``location field'' on that space, giving the spacetime properties (in the first instance: the locations) of the various parts of the body. The move from a global to a local theory then proceeds in exactly the same way, mathematically, for spacetime as for internal symmetries; in this sense at least, gravity turns out to be fully a gauge theory.

Many aspects of this framework have been discussed previously. The standard mathematical treatments of relativistic particles and strings in a background spacetime proceed in this way: the base manifold is a (1- or 2-dimensional) bare manifold and its location \emph{in} spacetime is a map from that base manifold \emph{to} spacetime. Treating a four-dimensional field theory this way (again, with a fixed background spacetime) has been explored as a toy model for quantum gravity, notably in discussions of the problem of time and the nature of general covariance.\footnote{The parametrised approach was first worked out in \cite{ishamkuchar1,ishamkuchar2}; for further development see, \egc, \cite{varadarajan-parametrized}.} In \cite{grignani} 
the gauge principle is applied to the parametrised motion of a particle (cf also \cite{stroblcomment,grignanistroblreply}) while in discussions of Poincar\'{e} gauge theory it is known that additional fields can be added to the theory to ameliorate its differences from Yang-Mills theory, and these additional fields can be identified \cite{hehlmetricaffine,gronwaldhehl} as the gauge versions of the spacetime-location field discussed above. So far as I am aware, however, there has not so far been a unified account of these various strands.

The structure of the paper is as follows. In sections \ref{gaugereview}--\ref{notgauge} I briefly review the gauge principle as applied to internal symmetries and the reasons why theories with local spacetime symmetry in their normal formulations fail to be gauge theories in this strong sense. In section \ref{parametrised} I give an account of the parametrised formalism applicable to global spacetime theories (in particular special relativity) and note some conceptual advantages of that formalism. In section \ref{gaugingpurelocation} I apply the gauge principle to the simplest of such theories: ``pure location theory'', the trivial theory where the field has only locational properties; the resultant theory, kinematically speaking, is general relativity. In section \ref{vectors} I generalise this analysis to apply to matter fields, including vector, tensor and spinor fields, and in section \ref{generalise} I briefly consider some possible generalisations. Section \ref{conclusion} is the conclusion.

\section{A review of the gauge principle}\label{gaugereview}

A (global, \iec non-gauge) field theory can be understood as a function \be \varphi:\mc{B}\rightarrow\mc{V},\ee where \mc{B} is the \emph{base space} (usually physical spacetime) and \mc{V} is the \emph{target space} whose points are the possible values of the field.\footnote{This restriction to scalar-valued fields (that is, excluding vectors or tensors on \mc{B}) may seem overly restrictive; I will show in section \ref{vectors} that it is no real restriction.} For maximum generality I will assume $\mc{B}$ is a differentiable manifold (possibly with additional structure, \eg a metric) and that $\mc{V}$ is a quite arbitrary mathematical space. In physics the most common choice is for $\mc{V}$ to be a vector space of some description (perhaps equipped with an inner product or other such structure), or else the real or complex field; for illustrative purposes we will often use $\mc{V}_N$, the $N$-dimensional real vector space equipped with a positive-definite inner product.

In most interesting cases, \mc{V} has a non-trivial group  of \emph{automorphisms}, \iec one-to-one mappings that preserve its mathematical structure. The automorphisms of a vector space are the invertible linear maps; the automorphisms of $\mc{V}_N$ are the orthogonal maps, etc. We can usefully distinguish between the abstract group and its representation $\mc{R}$ on \mc{V} --- so that the automorphism group of $\mc{V}_N$, for instance, is the abstract group $O(N,\mathrm{R})$, which is represented on the space by orthogonal linear maps. The automorphisms represent the \emph{internal symmetries} of the field theory. (In the case where there is a dynamical internal symmetry of the theory --- \iec a transformation of \mc{V} that preserves solutions of the equations --- that is not an automorphism  of \mc{V}, it is generally a sign that \mc{V} has surplus structure and can be simplified;\footnote{For instance, ``complex fields'' are often represented as maps to the complex plane, but since those theories normally have a phase symmetry, the real/imaginary distinction does no physical work, and the target space of the complex field is better taken as a one-dimensional complex vector space.} we shall suppose this done.)
We will largely restrict our attention to the group $\mc{G}_\mc{V}$ of \emph{small} automorphisms: that is, those that can be continuously deformed to the identity. (The large automorphisms present interesting issues of their own, but issues that lie beyond the scope of this article.) 

Given a field value $\varphi(x)\in \mc{V}$, we can distinguish between its \emph{absolute} features, which are invariant under the automorphisms, and its \emph{relative} features, which are not invariant. For $\varphi(x)\in\mc{V}_N$, the absolute features are determined by the magnitude $|\varphi(x)|$; facts about the direction of $\varphi(x)$ in $\mc{V}_N$ are relative, since they can be changed arbitrarily by the action of $\mc{G}_{V_N}$. Physically, the absolute features represent facts purely about the field at $x$, whereas the relative features represent facts about the field at $x$ compared to its value at other points: in the case of $\varphi(x)\in\mc{V}_N$, it makes sense to speak of (and in principle measure) $|\varphi(x)|$ intrinsically, whereas the angle of $\varphi(x)$ can only be coherently discussed, or measured, relative to a coordinate system defined by the field elsewhere in space. (Note that in the physics literature global symmetry transformations are typically described as occurring relative to some external reference frame not itself acted on by the symmetry.)

(One form of) the gauge argument begins with disquiet that the non-absolute features of $\psi(x)$ and $\psi(y)$ can be \emph{directly} compared, no matter the spatiotemporal separation of $x$ and $y$; as \cite{kiritsis} puts it, this seems ``at odds with  the `spirit' of relativity''. Leaving aside the conceptual justification of this move, what makes it possible \emph{mathematically} is that $\psi(x)$ and $\psi(y)$ lie in the \emph{same} mathematical space $\mc{V}$, and so we can remove this automatic comparability by equipping each point $x\in \mc{B}$ with \emph{its own} copy $\mc{V}_x$ of $\mc{V}$. A field  is no longer a function $\varphi$ from $\mc{B}$ to $\mc{V}$, but an assignment to each $x\in\mc{B}$ of a point $\tilde{\varphi}(x)\in \mc{V}_x$. Mathematically, this makes $\tilde{\varphi}$ a section of a $\mc{V}-$fibre bundle over $\mc{B}$ with structure group $\mc{G}_\mc{V}$.\footnote{To be more precise: the bundle here is an associated bundle over $\mc{B}$, with typical fibre \mc{V}, defined by some principal $\mc{G}_\mc{V}$-bundle over \mc{B}; there may, of courses, be many such principal bundles.}

Formulating our field theory \emph{just} as a section of a fibre bundle throws away \emph{too much} structure: we need to restore the ability to at least compare field values at infinitesimally close points $x,x+\delta x$, and we do so by introducing a $\mc{G}_\mc{V}$ \emph{connection} on the bundle, which allows us to parallel-transport a field value in $\mc{V}_x$ along a path to $\mc{V}_y$ so as to compare it directly to another such field value in $\mc{V}_y$. (Its precise mathematical formulation in the current rather abstract framework will not be needed; cf, \egc, \cite{kobayashinomizu}.) There is no conceptual requirement that the parallel transport rule is independent of the path taken from $x$ to $y$ (to impose such a requirement makes sense only if we have not abandoned the idea that direct comparisons of field values at distant points are permitted), so the connection in general introduces additional degrees of freedom into the theory. A theory of this kind, consisting of a field defined on a fibre bundle and a connection on that fibre bundle, is a \emph{local} (or \emph{gauge}) field theory.

The fibre-bundle formulation is elegant, but calculationally awkward compared to the formulation of global fields. At least locally,\footnote{Whether it can be done globally depends on the topology of the bundle} we can return to the global formalism by a choice of \emph{gauge}: a smooth choice for each $\mc{V}_x$ of an automorphism $f_x:\mc{V}_x\rightarrow \mc{V}$ to (a fixed copy of) $\mc{V}$. The field $\tilde{\varphi}(x)\in \mc{V}_x$ can now be represented by $\varphi(x)=f_x \cdot \tilde{\varphi}(x) \in \mc{V}$. The choice of gauge also fixes a representation of the connection as a \emph{covariant derivative} $\mc{D}_\mu$ on $\mc{B}\times \mc{V}$:
\be 
\mc{D}_\mu \varphi = \partial_\mu \varphi + A_\mu \cdot \varphi,
\ee
where $A_\mu$ is a one-form taking values in the Lie algebra $\mathrm{g}_\mc{V}$ of $\mc{G}_\mc{V}$, and $\cdot$ is the Lie algebra representation of $\mathrm{g}_\mc{V}$ induced by the representation of $\mc{G}_\mc{V}$ on $\mc{V}$. The field theory is then jointly represented, in this gauge, by $\varphi$ and $A_\mu$. In the case of $\mc{V}_N$, the field is now (kinematically) a Yang-Mills theory with structure group $SO(N)$ (or $O(2)$ for $N=2$).

The choice of gauge, however, is purely conventional. Any \emph{gauge transformation} between 
two gauges can be represented by a map $\Lambda:\mc{B}\rightarrow \mc{G}_V$, and transforms $\varphi$ and $A_\mu$ as follows:
\be \label{gaugetransform1}
\varphi(x) \rightarrow \Lambda(x)\cdot \varphi(x)
\ee
\be \label{gaugetransform2}
 A_\mu(x) \rightarrow \mathrm{Ad}(\Lambda(x))A_\mu(x) - \partial_\mu \Lambda(x)\Lambda^{-1}(x)
\ee
where $\mathrm{Ad}$ is the adjoint action and $\Lambda^{-1}$ acts on $\partial_\mu \Lambda$ by right translation (for matrix groups, $\mathrm{Ad}(X)Y=XYX^{-1}$ and the action of $\Lambda^{-1}$ is ordinary multiplication). We can then calculate that
\be \label{gaugetransform3}
\mc{D}_\mu \varphi(x) \rightarrow \Lambda(x) \cdot \mc{D}_\mu\varphi(x)
\ee

Notice that these \emph{gauge symmetries} are purely internal, acting (independently) on each fibre and leaving the points of the base manifold \mc{B} invariant. The full symmetry group of the theory will be the gauge symmetries \emph{together with} any symmetries of \mc{B} itself.

In calculations, it can be useful to eliminate this gauge freedom by \emph{gauge-fixing}: picking, for each equivalence class of $(\varphi,A_\mu)$ pairs under gauge transformations, a single preferred member. Examples in $\mc{V}_N$ are the \emph{Lorenz gauge}  in which we require\footnote{There is a background metric assumed here.} that $\partial^\mu A_\mu=0$ and that $A_\mu$ vanishes at spatial infinity, and the \emph{unitary gauge} in which we pick a particular direction in $\mc{V}$ and require $\varphi(x)$ to lie in that direction, so that the only residual degree of freedom of $\varphi$ is $|\varphi|$.

Interpretationally, the field theory represents (a) the absolute features of the field at spacetime points, and (b) the path-dependent relations between values of the field at distinct points. These features are represented \emph{jointly and redundantly} by $\varphi$ and $A_\mu$, but there is no general division of labour: it is not the case, in particular, that $\varphi$ represents (a) and $A_\mu$ represents (b), though this may be true in a particular choice of gauge-fixing (it is true for $\mc{V}_N$ in the unitary gauge). In many cases (specifically, when $\mc{G}_\mc{V}$ acts freely on $\mc{V}$, perhaps except for some subset of \mc{V} of measure zero) specification of $\rho$ and $\mc{D}_\mu\rho$ (that is, specification of the absolute features of the field and of the relations between \emph{infinitesimally close} points) suffices in general to allow us to solve for $\varphi$ and $A_\mu$ up to gauge transformations, and thus to specify the entire theory.\footnote{In Yang-Mills theory, this can be seen by adopting the unitary gauge(\cite{Weinberg1971}; see, \egc, \cite{peskinschroeder} for details), which is defined only for $\varphi \neq 0$.}

Before continuing I should note that the gauge argument does not uniquely fix the \emph{dynamics} of the newly local theory, even given the dynamics of the global theory; however, insofar as classical field theory is of interest mainly as a route to quantum theory this is of comparatively little significance given that renormalisation considerations generally fix the form of a theory's Lagrangian almost uniquely (\iec, up to finitely many experimentally determined parameters) once the theory's symmetries are known. Throughout this paper I will be content to recover the kinematic form of a theory, leaving its dynamics to be determined by symmetry and renormalisation considerations.

\section{Why (normal formulations of) general relativity are not gauge theories (in this sense)}\label{notgauge}

General relativity (GR) is often introduced by starting with Minkowski spacetime, representing it in differential-geometric form via a metric and an affine connection, and then dropping the requirement that the metric and connection are curvature-free. There is a clear family resemblance to the gauge argument, and much ink has been spilled trying to understand GR in this way: it has been considered, variously, as a gauge theory of the diffeomorphism group, the general linear group $\mathrm{GL}(4,\mathbf{R})$, the translation group $\mathrm{R}^4$, the affine group of linear maps and translations (formally $\mathrm{R}^4 \rtimes \mathrm{GL}(4,\mathrm{R})$), the Lorentz group $\mathrm{SO}(3,1)$, the Poincar\'{e} group of Lorentz boosts and translations ($\mc{P}=\mathrm{R}^4 \rtimes \mathrm{SO}(3,1))$, and probably others besides. These various theories undoubtedly have much in common with `standard' gauge theories, and whether they are `real' gauge theories can quickly become an exercise in semantics.

It is, however, possible without semantic tangles to ask and answer the question of whether GR, in a given formulation (and with or without matter), is a gauge theory in the specific sense of section~\ref{gaugereview}: that is, whether it is a field theory specified by a section of some $\mc{V}-$bundle over a base manifold $\mc{B}$ together with a connection on that bundle, or (equivalently) whether it is specified, for some $\mc{V}$, by a function $\varphi:\mc{B}\rightarrow
\mc{V}$ and a one-form $A_\mu$ taking values in $\mathrm{g}_\mc{V}$, with (\ref{gaugetransform1}--\ref{gaugetransform2}) as symmetries. In at least the standard formulations, the answer is `no', and the reasons are instructive. (For simplicity, henceforth `gauge theory' means exclusively `gauge theory in \emph{my} sense.)

The \textbf{traditional formulations} of GR regard it as the theory of a metric tensor field $g_{\mu\nu}$ and a covariant derivative $\nabla_\mu V^\nu=\partial_\mu V^\nu + \Gamma^\nu_{\mu\tau}V^\tau.$ GR proper requires that $\nabla_\tau g_{\mu\nu}=0$, and $\Gamma^\mu_{[\nu\tau]}=0$ (`torsion freedom') which suffices to fix $\nabla$. \textbf{Einstein-Cartan theory} (or Einstein-Cartan-Sciama-Kibble theory) is obtained by dropping the torsion-freedom requirement, and \textbf{metric-affine gravity} by additionally dropping the assumption of metric compatibility.\footnote{For the traditional formalisms see, \egc, \cite{mtw} or \cite{waldrelativitybook}; for the generalisations, see \cite{hehlblagojevicbook} and references therein.}

These fail to be gauge theories for several reasons. Most significantly (and even if $g$ is set aside entirely) $\nabla_\tau$, while a covariant derivative of \emph{some} kind, is not a covariant derivative in the sense of section \ref{gaugereview}, since it is defined on the \emph{tangent} bundle $T\mc{B}$, and this bundle has more structure than the fibre bundles previously considered.

To elaborate, we \emph{can} think of $T\mc{B}$ as a vector bundle, with $T_x\mc{B}$ a 4-dimensional vector space. But there is a preferred identification between \emph{vectors} in the fibre at $x$ and infinitesimal paths in $\mc{B}$ through $x$ --- indeed, in standard formulations of the tangent bundle the vectors in $T_x\mc{B}$ are \emph{identified} with those infinitesimal paths, or equivalently (if mathematically more conveniently) with derivative operators along those paths. It is this relation between the base manifold and the tangent bundle that makes it possible to regard the derivative $\dot{\gamma}(\lambda)$ of a path $\gamma(\lambda)$ through \mc{B} as lying in $T_{\gamma(\lambda)}\mc{B}$. 

This shows up in the mathematical form of the connection coefficients $\Gamma^\mu_{\nu\tau}$. This \emph{can} be thought of as a $GL(4,\mathrm{R})$-valued 1-form (with the $\mu$,$\tau$ coefficients interpreted as $GL(4,\mathrm{R})$ indices and $\nu$ as a 1-form index). But the indices really index the same space, so that in particular it makes sense to impose torsion-freedom, a requirement with no analogue for internal connections (since there the lower indices of the connection index coordinates in different spaces). 

It shows up also in the automorphisms of the bundle, and hence in the symmetries of the theory. Whereas in section \ref{gaugereview} there was a sharp separation between symmetries of the base manifold, and internal symmetries that leave base-manifold points unchanged, in this case there are \emph{no} purely internal symmetries, since any nontrivial transformation of $T_x\mc{B}$ will break the relation between vectors and infinitesimal curves. The only automorphisms of the tangent bundle are those induced by diffeomorphisms of the base manifold.

(The literature is inconstant as to whether the tangent bundle should be called a fibre bundle or not; for clarity, henceforth I call fibre bundles in the original sense \emph{internal} fibre bundles.)

More flatfootedly, the metric field $g_{\mu\nu}$ is neither a section $\varphi$ of an internal fibre bundle, nor a connection of any kind. (It can be understood as a section of the $(0,2)$ tensor bundle over $\mc{B}$, but this, like the tangent bundle, is not an internal fibre bundle).

The \textbf{tetrad formalism} of GR\footnote{Reviewed in, \egc, chapter 2 of \cite{rovelli}.} starts with the 4-dimensional Minkowski vector space $\mc{M}$, equipped with metric $\eta_{ab}$, and with the Lorentz group $\mathrm{SO}(3,1)$ as its automorphism group. (I use Roman letters here as indices in \mc{M}.) The theory is formulated on a fibre bundle over \mc{B} with typical fibre $\mc{M}$, and so looks rather more like a gauge theory; it is specified (again setting aside matter fields) by a \emph{tetrad} $e$ (a one-form taking values in the fibres, \iec $e(x):T_x\mc{B}\rightarrow \mc{M}_x\mc{B}$) and a covariant derivative on $\mc{M}\mc{B}$. With respect to a particular gauge (and using Roman indices to index vectors in \mc{M}) the tetrad is represented by a 1-form $e^a$ taking values in \mc{M}, and the covariant derivative by a $\mathrm{SO}(3,1)$-valued 1-form $\omega^a_b$, as in
\be 
\nabla_\mu = \partial_\mu + \omega^a_{b\mu}.
\ee
From here there are several ways to connect connection and tetrad. \textbf{Conventional GR in tetrad form} is reproduced by imposing the \emph{torsion-freedom} constraint 
\be T[e,\omega] = \mathrm{d}e^a + \omega^a_b \wedge e^b=0,\ee
which determines $\omega$ uniquely. \textbf{Teleparallel gravity}\footnote{\cite{Hayashi1967,Pellegrini1963}; see \cite{andrade} for a brief review, \cite{aldrovandi} for a book-length treatment.).} is reproduced by instead imposing the condition $\omega=0$, so that the connection has no curvature in the usual sense ($\omega=0$ is not invariant under local Lorentz transformations, so this is not an invariant specification.)
In the more general case of \textbf{Poincar\'{e} gauge theory}\footnote{x} the tetrad and the connection are treated as independent fields.

$\omega$, by itself, is an $SO(3,1)$ connection in the full sense of section \ref{gaugereview}, with exactly the expected transformation properties: in particular, the theory has, as expected, local $SO(3,1)$ transformations as symmetries. Furthermore, these remain symmetries even when $e^a$ is included: if $\Lambda:\mc{B}\rightarrow SO(3,1)$ is a gauge transformation, the overall symmetry is
\be 
e^a_\mu(x) \rightarrow \Lambda^a_b(x) e^b_\mu(x) 
\ee
\be 
\omega^a_{b\mu} (x) \rightarrow \Lambda^a_c(x)\omega^c_{d\mu}(x)(\Lambda^{-1})^d_b(x) - \partial_\mu \Lambda^a_b(x).
\ee
But the $e^a$ field spoils tetrad gravity's pretensions to be an $\mathrm{SO}(3,1)$ gauge theory in 
our sense. It is not a $\mc{M}$-valued \emph{function} but an $\mc{M}$-valued \emph{one-form}, and so cannot be understood as a section of the $\mc{M}$-bundle. In fact, it has a rather different natural geometric interpretation which belies the interpretation of that bundle as an \emph{internal} bundle: for each $x$ it gives a 1-1 map from $T_x\mc{B}$ to $\mc{M}_x \mc{B}$, allowing us to identify the Minkowski vector bundle with the tangent bundle. From this perspective the $\mathrm{SO}(3,1)$ freedom rather mundanely represents our freedom to coordinatise the tangent bundle as we like.

An alternative is to interpret $e^a$ as another connection: the translational part of the Poincar\'{e} group \mc{P}. (This is the main conceptual idea behind teleparallel gravity.) From this perspective $(\omega,e)$ is a one-form taking values in the Lie algebra of \mc{P}. Suggestively, if we formally calculate the curvature under this interpretation we get the $\mathrm{p}$-valued 2-form $(R,T)$, where $R$ is the curvature of the $SO(3,1)$ connection and $T$ is the torsion. However, the theory still cannot be a gauge theory in \emph{our} sense: the fibre  $\mc{M}$ is a vector space, and does not bear any representation of \mc{P}. We can attempt to interpret the translations as acting on the base manifold \mc{B} instead, taking a point $x^\mu$ to $x^\mu+\delta x^\mu$ (this is the standard route used to arise at Poincar\'{e} gauge theory), but invariance under this transformation is just infinitesimal diffeomorphism invariance, and does not have the purely internal form we require for a gauge symmetry).

These various theories of gravity have in common that they represent gravitation and curvature on the base space of the theory, and hence make use of the tangent bundle. In the next section I will show how the concept of parametrisation lets us put gravitational physics on a par with other gauge theories, by moving the geometry entirely off the base space.

\section{Parametrised field theory}\label{parametrised}

Field theories as standardly conceived are assignments of quantities to spacetime points, and as such differ sharply from particles or strings, which are extended bodies parts of which occupy spacetime points. A particle in special relativity is represented as a map from its one-dimensional worldline to Minkowski spacetime; a string, as a map from its two-dimensional world\emph{sheet} to spacetime. In each case, the underlying worldline or worldsheet is just a bare differentiable manifold, and the theory therefore has two distinct sets of symmetries: a diffeomorphism symmetry of the underlying manifold, and the global symmetries of the spacetime. In fact, particles and strings are global field theories in the sense of section \ref{gaugereview}, being maps from a 1- or 2-dimensional underlying space to a target space which is just Minkowski spacetime. 

We should pause a moment to be clear what ``Minkowski spacetime'' actually means here. It had better not be the Minkowski \emph{vector space} \mc{M}, which has a preferred origin; however, it is overkill to take it to be the Minkowski \emph{manifold}, equipped with a flat metric tensor field and associated connection: the differential-geometric machinery used is mostly only necessary for curved spacetimes. The mathematically simplest option is to treat it as an \emph{affine space}: a set of spacetime points $\mc{A}\mc{M}$, together with a rule assigning to each pair $x,y$ of points in $\mc{A}\mc{M}$ a vector $(y-x)$ in the Minkowski vector space $\mc{M}$, such that $(x-x)=0$ and $(z-y)+(y-x)$ = $z-x$.\footnote{More formally, affine Minkowski space is a set together with a free transitive action on that set of \mc{M} regarded as an additive group.} The automorphism group of $\mc{A}\mc{M}$ is, as we might have hoped, the Poincar\'{e} group \mc{P}.

Since the particle (or string) is already a global field theory with the right automorphism group, we could proceed directly to seeking out a gauged version of it (this is the strategy of \cite{grignani}). However, to apply the gauge argument directly we need something more like a \emph{conventional} field theory, which is to say that we need a way to write ordinary field theories in something like the particle/string form.

This can be achieved via \emph{parametrised field theory}. Given a field $\bar{\varphi}:\mc{A}\mc{M}\rightarrow \mc{V}$ on Minkowski spacetime with some Lagrangian $L(\bar{\varphi},\partial_a\bar{\varphi})$, and a bare manifold $\mc{M}$ diffeomorphic to $\mathbf{R}^4$,we can define its parametrised version as a field
\be 
\varphi:\mc{B} \rightarrow \mc{A}\mc{M}\oplus \mc{V};\,\,\,\, \varphi(x)=(\rho^a(x),\psi(x))
\ee
with Lagrangian
\be 
\mc{L}(\rho,\partial_\mu \rho,\psi,\partial_\mu\psi)=L(\psi,e^\mu_a \partial_\mu\psi)\mathrm{det}(e^a_\mu)
\ee
where $e^a_\mu=\partial_\mu \rho^a$, and $e^\mu_a$ is the inverse of $e^a_\mu$, \ie $e^a_\mu e^\mu_b=\delta^a_b.$ In the case of a scalar field with Lagrangian
\be 
L(\psi,\partial_a \psi)= \frac{1}{2}\eta^{ab} \partial_a \psi \partial_b \psi - V(\psi),
\ee
for instance, the parametrised version has Lagrangian
\be  
\mc{L}(\rho,\partial_\mu \rho,\psi,\partial_\mu\psi)= 
\left(\frac{1}{2}\eta^{ab}e^\mu_a e^\nu_b \partial_\mu \psi \partial_\nu \psi + V(\psi)\right)
\mathrm{det}(e^a_\mu)
\ee
and writing $g_{\mu\nu}=\eta^{ab}e_\mu^ae_\nu^b$ lets us reexpress this in the more familiar form
\be 
\mc{L}(g_{\mu\nu},\psi,\partial_\mu\psi)=
\left(\frac{1}{2}g^{\mu\nu}\partial_\mu \psi \partial_\nu \psi - V(\psi)\right) \sqrt{\mathrm{det}g}.
\ee
$\rho^a$ is the \emph{location field}, which says where in spacetime any given part of the body manifold is.

Parametrisation has mostly been seen as a rather cheap way to make a theory generally covariant  and discussed mostly as a toy theory for quantum gravity \cite{ishamkuchar1,ishamkuchar2,varadarajan-parametrized}. However, it has some virtues conceptually even aside from the gauge argument:
\begin{itemize}
\item It is most naturally thought of as representing a field as an extended body whose parts occupy various spacetime points and have various non-spatiotemporal properties. As such, it treats spacetime and internal degrees of freedom very much on a par: Minkowski spacetime and the `real' internal space $\mc{V}$ both parametrise properties that given parts of the field might have.
\item It provides for a rather clean separation between the substantive symmetries of the theory (which are represented by the automorphisms of spacetime $\mc{A}\mc{M}$ and of the internal space \mc{V}) and the diffeomorphism symmetry, which can be made a symmetry of any spacetime theory through appropriate reformulation. We have made the theory generally covariant, but without reducing the spacetime symmetries to special cases of the diffeomorphism symmetries. 
\item It makes for a cleaner understanding of Noether's theorem, since the distinction between ``dependent'' and ``independent'' variables is eliminated: all variables are dependent variables, even the spacetime ones.\footnote{If desired we can treat the diffeomorphism symmetry as an independent-variable transformation --- but the Noether currents associated to that symmetry vanish identically in any case.} For instance, the infinitesimal translation symmetry of a parametrised scalar field theory $\varphi=(\rho^a,\psi)$ is just
\be 
\rho^a\rightarrow \rho^a + \xi^a;\,\,\,\,\, \psi \rightarrow \psi
\ee 
and the associated Noether current is
\be 
J^{a\mu} = \pbp{\mc{L}}{e_\mu^a}.
\ee
Assuming that $\mc{L}=L \mathrm{det}(e^a_\mu)$ and that $L$ depends on $\rho$ only through $g_{\mu\nu}$, this yields the familiar-looking
\be 
J^{a\mu}= e_\nu^a\left(\pbp{L}{g_{\mu\nu}}  + \frac{1}{2}g^{\mu\nu} L \right)\sqrt{\mathrm{det}g}
\ee
\item It allows us to treat vector, tensor and spinor fields (in special relativity) straightforwardly as functions, as we will see in section \ref{vectors}.
\end{itemize}

\section{Gauging pure location theory}\label{gaugingpurelocation}

The simplest parametrised field theory is the trivial `pure location theory', where the \emph{only} field is the location field $\rho:\mc{B}\rightarrow \mc{A}\mc{M}$. Under the additional requirement that $\rho$ is a smooth $1:1$ mapping, this is just empty Minkowski spacetime in parametrised form; however, the gauge recipe of section \ref{gaugereview} still applies directly to it and yields a highly non-trivial theory (at least kinematically; I continue to leave the dynamics unspecified). The details are as follows: 
\begin{enumerate}
\item We localise the theory by replacing $\rho$ with a section $\tilde{\rho}$ of a bundle over $\mc{B}$. The resultant bundle $\mc{AMB}$ has typical fibre $\mc{AM}$: it is an \emph{affine bundle}, rather than the vector bundles more familiar from Yang-Mills theory. It is still, however, an internal bundle, with no particular connection to the tangent bundle $T\mc{B}$.
\item We introduce a connection on $\mc{AMB}$ to allow us to compare the values of the location field $\tilde\rho$ at infinitesimally close points. With respect to a choice of gauge (that is, a smooth choice for every $x$ of a map  $f_x:\mc{AMB}\rightarrow\mc{AM}$, $\tilde{\rho}$ is represented by a function $\rho:\mc{B}\rightarrow \mc{AM}$ and the connection by the covariant derivative
\be 
\nabla_\mu \rho^a = \partial_\mu \rho^a + (A_\mu \cdot \rho)^a,
\ee
defined in terms of a one-form $A_\mu$ taking values in the Lie algebra of the Poincar\'{e} group.
\end{enumerate}
Given some arbitrary choice $0$ of origin of $\mc{A}\mc{M}$, we can decompose any Poincar\'{e} group element into a rotation around $0$ and a translation. $A_\mu$ can then be broken into infinitesimal rotational and translational parts $(\omega^a_{\mu b},\tau^a)$, and the partial derivative further decomposed as
\be 
\nabla_\mu \rho^a = \partial_\mu \rho^a + \omega^a_{\mu b} \rho^b + \theta^a.
\ee
A gauge transformation is a  map $\Lambda:\mc{B}\rightarrow\mc{P}$, which we can decompose as $\Lambda(x)=(R^a_b(x),\xi^a(x))$, and its action on the location field and the connection can be read off from (\ref{gaugetransform1}-\ref{gaugetransform3}), or else verified explicitly (for readability I use matrix notation and suppress the explicit Minkowski indices):
\be 
\rho(x) \rightarrow R(x)\rho(x)+ \xi(x)
\ee
\be 
\omega_\mu(x) \rightarrow \mathrm{Ad}(R(x))\omega_{\mu}(x)  - \partial_\mu R(x)R^{-1}(x)
\ee
\[ 
\theta(x) \rightarrow R(x)\theta(x) - \mathrm{Ad}(R(x))\omega_{\mu}(x)\xi(x)
\]
\be 
 -\partial_\mu\xi(x)+ \partial_\mu R(x)R^{-1}(x)\xi(x)
\ee
\be \label{covderiv-location}
\nabla_\mu\rho(x) \rightarrow R(x)\nabla_\mu\rho(x).
\ee
(If you were expecting a translation term in (\ref{covderiv-location}), note that the derivative of a function to $\mc{AM}$ takes values in the tangent space of $\mc{AM}$, \iec the Minkowski vector space $\mc{M}$, and the translational part of \mc{P} acts trivially on $\mc{M}$. In the case of Yang-Mill theories where the target space is a vector space, this subtlety does not arise since a vector space is its own tangent space.)

This \emph{local pure location theory} is a gauge theory in the full sense of section~\ref{gaugereview}, defined on an internal bundle and with ``matter'' field and connection transforming exactly as the gauge recipe stipulates. It has the full local Poincar\'e group (the group whose elements are maps from the base space to the Poincar\'{e} group) as a symmetry group, in addition to another symmetry group generated by the diffeomorphisms on the body manifold \mc{B}, and the two have conceptually entirely different status: the former follows from the global Poincar\'{e} symmetry on the fibres and the imposition of locality, the second from the fact that points on the body manifold have no absolute properties of their own and so can be smoothly permuted without affecting the physics.  

The curvature of the connection is as usual a Poincar\'{e}-valued 2-form, and can be decomposed into its rotational and translational parts $R^a_{b\mu\nu}$,$T^a_{\mu\nu}$: in differential-form notation, it can be evaluated explicitly as
\be 
\mc{R}^a_b= d \omega^a_b + \omega^a_c \wedge \omega^c_b
\ee
\be 
\mc{T}^a_b = d \theta^a + \omega^a_b \wedge \theta^b.
\ee
Since the Poincar\'{e} group acts transitively on \mc{AM}, the location field $\rho(x)$ has no meaning by itself: any two locations are intrinsically identical. (This was also true before gauging the theory.) What has physical significance is the covariant derivative, which represents the displacement between two infinitesimally close points on the body manifold. Integrating this displacement over finite paths in the usual way for gauge theories (\iec, keeping track of noncommutativity) will give the displacements between finitely separated points, but now that the theory has been localised the displacement will in general be path dependent. The curvature then gives the \emph{infinitesimal anholonomy} of the location field: the net displacement between a point and itself around an infinitesimal curved path:
\be 
T^a_{\mu\nu} = \mc{R}^a_{b\mu\nu}\rho^a + \mc{T}^a_{\mu\nu}.
\ee 
 (It does so rather redundantly --- ten components of curvature provide only a four-component displacement --- reflecting the fact that the Poincar\'{e} group does not act \emph{freely} on Minkowski space: given any two points, there are a large number of different Poincar\'{e} group actions  that map one to the other. When we consider more complicated fields in section~\ref{vectors}, this redundancy will be eliminated.) Anticipating later results, we call this infinitesimal anholonomy the \emph{torsion}. Note that it cannot be identified simply with the translational part of the curvature: it contains both translational and rotational parts.)
 
We can connect pure location theory to familiar physics by a choice of gauge. Specifically, let us adopt the \emph{stationary gauge}, in which $\rho(x)=0$ (that is, equals the arbitrarily chosen `origin' of \mc{AM}) for all $x$.\footnote{\cite{grignani} call it the `physical' gauge; I avoid this terminology since from a gauge-theory perspective, no gauge is more physical than any other.} 
This is not a complete fixing of gauge: it breaks the local Poincar\'{e} symmetry down to the local Lorentz group, since any Lorentz transformation leaves the origin invariant. (There is an analogy with the breaking of $SU(N)$ down to $U(1)$ in internal symmetry breaking: $U(1)$ is the subgroup of $SU(N)$ under which the (arbitrarily-selected) lowest-energy state is invariant.)

In the stationary gauge, we have simply
\be 
\nabla_\mu \rho^a = \theta^a_\mu
\ee
\be T^a_{\mu\nu}= (d \theta^a + \omega^a_b \wedge \theta^b)_{\mu\nu}:
\ee
in other words, \emph{in the stationary gauge} the translational part of the connection represents the 
torsion of the location field. If we write
\be 
e^a_\mu = \nabla_\mu \rho^a,
\ee
then since the rotational part of the connection is translationally invariant,
we obtain a translationally invariant expression for the torsion:
\be \label{torsion}
T^a = (d e^a + \omega^a_b \wedge e^b).
\ee
The covariant derivative $e$ and the rotational connection $\omega$ provide a complete specification of the theory up to gauge invariance, and can of course be recognised (kinematically) as the tetrad and connection of tetrad-formalism gravitation. Or to put it another way around: the location field allows us to interpret the tetrad as the covariant derivative of the location field, and the connection as the rotational part of the Poincar\'{e} connection for that field. The tetrad can be identified directly with the translational part of the connection \emph{only} in the stationary gauge (which is reflected in the well-known fact that it transforms homogenously under translations).

Similarly, expression (\ref{torsion}) can be recognised as the standard definition of the torsion. The parametrised formalism then illustrates in a very simple and direct way the interpretation of torsion as giving the infinitesimal translation around a closed loop.

The stationary gauge also allows us to deduce that pure location theory is specified entirely, up to gauge freedom, by the covariant derivative and the torsion. For (\ref{torsion}) can be solved for $\omega$, yielding the full connection in the stationary gauge (and of course $\rho$ is trivially known in that gauge). In the case where the torsion is constrained to vanish, we (kinematically) recover general relativity. Conversely, by requiring the rotational part of the connection to vanish ($\omega=0$) we can express the theory entirely in terms of $e$. Such a theory is a gauge theory of pure translation: since the translation group acts freely on $\mc{AM}$ we would expect the covariant derivative of $\rho$ to specify the complete theory, and indeed it does: torsion is given gauge-invariantly by
\be  
T^a = \mathrm{d} (\nabla \rho^a)\equiv \mathrm{d} e^a.
\ee
This theory can be recognised as teleparallel gravity (note that the $\omega=0$ recipe is not invariant under local Lorentz invariance, reflecting the well-known Lorentz-dependence of the Weitzenbock connection of teleparallelism).

\section{Vector, tensor, and spinor fields}\label{vectors}

A vector field on Minkowski spacetime can be treated as a map $\varphi=(\rho,\psi)$ where $\rho$ takes values in Minkowski spacetime $\mc{A}\mc{M}$ and $\psi$ in the Minkowski vector space $\mc{M}$. The map assigning an element ($x-y$) of $\mc{M}$ to any $x,y\in \mc{A}\mc{M}$ serves to `solder' the two together and distinguishes the vector field from a purely internal vector field. Conceptually, each part of the body manifold has both a location in spacetime and a vector field value.

It will be convenient to regard vectors as living in their own copy of $\mc{M}$, equipped with a preferred identification with the copy of \mc{M} used to define $\mc{A}\mc{M}$, so that parametrised fields take values in $\mc{A}\mc{M}\times \mc{M}$; similarly, tensor fields take values in the tensor products of Minkowski vector space equipped with such a preferred identification. I write $\mc{A}\mc{M}\bar{\times} \mc{V}$ for any vector or tensor space so linked to Minkowski space.

This framework is extremely similar to the way we represented scalar fields with internal degrees of freedom: in both cases the field is a map associating to each point of the base manifold both a location in $\mc{AM}$ and a field strength in some vector space. The difference is that the preferred identification means that the automorphism group of $\mc{AM} \bar{\times} \mc{B}$ is not simply the product of the automorphism groups of $\mc{AM}$ and $\mc{B}$: rather, the automorphism group is the Poincar\'{e} group, with the translational part acting on \mc{AM} alone and the rotational part acting jointly on \mc{AM} and \mc{B}.

For spinors, we take $\mc{C}$ to be the 2-dimensional complex vector space, and fix a preferred bilinear map $\epsilon^a_{\alpha\alpha'}$ from $\mc{C}\otimes \mc{C}^*$ to $\mc{M}$. Parametrised spinor fields now take values in the space $\mc{A}\mc{M}\times \mc{C}$ equipped with this map; the automorphism group is now $\mathbf{R}^4 \rtimes \mathrm{SL}(2,\mathbf{C})$, the double covering of the Poincar\'{e} group. I extend the notation $\mc{A}\mc{M}\bar{\times} \mc{V}$ to include the spinor case also. 

Finally, for vector fields having additional internal degrees of freedom represented by some vector space $\mc{U}$, the field strengths take values in $\mc{M}\otimes \mc{U}$, with this copy of $\mc{M}$ again identified in a preferred way with the copy of \mc{M} used to define \mc{AM}; this extends naturally to spinor and tensor fields with internal degrees of freedom, and again I extend the $\bar{\times}$ notation to cover this case.

This somewhat abstract discussion is perhaps clearer in index form: a general tensor field on Minkowski spacetime is represented in the parametrised format by a location field $\rho^a$ together with some object \be \psi^{a_1\ldots a_n\alpha_1\ldots \alpha_k}_{b_1\ldots b_m\beta_1\ldots \beta_l}\ee
where $a_i,b_i$ are spacetime indices and $\alpha_i,\beta_i$ are internal-space indices. An infinitesimal internal symmetry acts on the $\alpha_i,\beta_i$ indices and leaves $\rho^a$ invariant; an infinitesimal translation acts on $\rho^a$ alone; an infinitesimal Lorentz transformation acts on the spacetime indices of $\psi$ and on $\rho^a$. For spinor fields, the $a_i$ and $b_i$ become spinor indices. Finally, the most general case of multiple such fields is represented by a tuple
\be 
(\rho,\psi_1,\ldots \psi_n)
\ee
where the $\psi_i$ are vector or spinor fields of the above form.

A global theory formulated in this way has as its symmetry group, as we would expect, the product of (i) the Poincar\'{e} group \mc{P}; (ii) the internal symmetry group $\mc{G}_{int}$, which acts trivially on \mc{AM}; (iii) the diffeomorphism group.

Even a theory as complicated as this is still ultimately a map from $\mc{B}$ to some space $\mc{V}$ on which some symmetry group $\mc{G}$ (in this case $\mc{P}\times \mc{G}_{int}$) acts, and so can still be localised via the gauge argument. This yields as usual a theory on a fibre bundle with typical fibre \mc{V}, equipped with a connection $A_\mu$ taking values in the direct sum of the Lie algebras of $\mc{P}$ and $\mc{G}_{int}$. This connection may be decomposed into a triple
\be 
A_\mu=(\theta_\mu,\omega_\mu,B_\mu)
\ee
where $B_\mu$ is a one-form taking values in $\mathrm{g}_{int}$ and $\theta_\mu$,$\omega_\mu$ are the translational and rotational parts of the Poincar\'{e} connection introduced in section \ref{gaugingpurelocation}. The covariant derivative defined by that connection acts in the standard way:
\be 
D_\mu (\rho,\psi_1,\ldots \psi_n)= \partial_\mu (\rho,\psi_1,\ldots \psi_n) + A_\mu \cdot (\rho,\psi_1\ldots \psi_n).
\ee
In the simple case of a single vector field $\psi^a$ with no internal degrees of freedom, for instance, we have
\be 
D_\mu (\rho^a,\psi^a) = (\partial_\mu \rho^a + \omega^a_{\mu b}\rho^b + \theta_\mu^a, \partial_\mu\psi^a + \omega^a_{\mu b}\psi^b) \equiv (e_\mu^a,D^L_\mu \psi^a)
\ee
where following section \ref{gaugingpurelocation} we continue to write 
\be 
e_\mu^a \equiv \partial_\mu \rho^a +  \omega^a_{\mu b}\rho^b + \theta_\mu^a
\ee
and we define
\be 
D^L_\mu \psi^a \equiv \partial_\mu \psi^a + \omega^a_{\mu b}\psi^b
\ee
as the \emph{Lorentzian covariant derivative} (it generalises in the obvious way to tensor and spinor fields). It follows from the general structure of gauge theory that this covariant derivative transforms via the tangent representation of $\mc{P}$ (that is, the Lorentzian part) under a gauge transformation $\Lambda(x)=(R^a_b(x),\xi^a(x))$: indeed,
\be 
(e_\mu^a,D^L_\mu \psi^a) \rightarrow (e_\mu^a,D^L_\mu \psi^a) + (R^a_b e_\mu^b,R^a_b\psi^b).
\ee
The gauge and location fields, meanwhile, transform just as in section \ref{gaugingpurelocation}, while $\psi$ transforms under the Lorentzian part of the gauge transformation. The theory, that is, is a gauge theory of the Poincar\'{e} group in a totally standard sense, with symmetry group being the semidirect product of the local Poincar\'{e} group and the diffeomorphism group.

If, however, we move to stationary gauge, we can eliminate $\rho$ entirely from the theory, and equate $e_\mu^a$ with the translational part of the gauge connection, the local Poincar\'{e} symmetry is broken down to a local Lorentz symmetry and the transformation laws become (suppressing indices)
\begin{eqnarray}
e_\mu & \rightarrow & e_\mu + R \cdot e_\mu \label{tetrad1} \\
\omega_\mu & \rightarrow & \omega_\mu + \mathrm{ad}(\omega_\mu)R - \partial_\mu R \\
\psi & \rightarrow & \psi + R\cdot \psi\label{tetrad3}
\end{eqnarray}
together with the usual action of the diffeomorphism group. This is the familiar symmetry group of tetrad gravity with dynamical connection: it is a hybrid theory, with some features of Lorentzian gauge theory but with the tetrad not identifiable straightforwardly either as a matter field or as a gauge connection. 

(Given the transformations (\ref{tetrad1}--\ref{tetrad3}) it is possible, conversely, to ask how the theory can be modified to allow $e_\mu$ to be interpreted as the translational part of a Poincar\'{e} connection. This can be done by introducing auxiliary fields\cite{hehlmetricaffine,gronwaldhehl}, which from the 
present viewpoint may be identified as the components of the location field.)

The full framework of location field, ordinary field strength, and translational and rotational connection is, of course, more complicated than the normal tetrad form, and contains a redundancy that for calculational purposes is no doubt usually best eliminated by moving to the stationary gauge. It is gratifying, however, to see that when the full framework is kept in view, our theory (at least kinematically) is simply, straightforwardly, a gauge theory of the Poincar\'{e} group, and interesting to note how, as elsewhere in physics, choice of a specific gauge obscures the symmetry structure even as it makes the theory's dynamical degrees of freedom more transparent.

Finally, note that the gauge theory we have constructed includes (or could include) two very different kinds of ``vector fields'':
\begin{enumerate}
\item Matter fields where the field's internal space includes a vector degree of freedom transforming under the Lorentz symmetry;
\item Gauge connections (Poincar\'{e} or internal) represented by one-forms.
\end{enumerate}
Conceptually the former is a property of a spacetime \emph{point}; the latter represents the infinitesimal relationship \emph{between} points. For the former, the ability to represent it as a \emph{tangent-space} object (via $V^a=V^\mu e_\mu^a$ is incidental and relies on an additional spatiotemporally extended object, the covariant derivative of the location field. For the latter, its (co)tangent nature is essential to its role in relating two different (infinitesimally separated) point.

It is therefore unsurprising (as well as following directly from our analysis) that the two sorts of vector field couple rather differently to the spacetime covariant derivative. In moving from flat to curved spacetime, ordinary derivatives of matter fields, including vector fields, get replaced by covariant derivatives, but (as is well known in the Poincar\'{e} gauge-theory
literature; cf the discussion in \cite{hehl-erice}) ordinary derivatives of gauge connections are not so replaced: indeed, the expression
\be 
\nabla_\mu A_\nu - \nabla_\nu A_\mu = (\partial_\mu A_\nu - \partial_\nu A_\mu) - T^\sigma_{\mu\nu}A_\sigma
\ee
is not gauge invariant unless the torsion vanishes. The gauge connection is itself \emph{part of} a unified covariant derivative, representing that part of the derivative concerned with internal-space relations.

\section{Possible generalisations}\label{generalise}

Richard Feynman once observed \cite[p.168]{feynmancharacter} that different formulations of the same theory, even if they are mathematically interchangeable, can point towards different generalisations to new physics. The parametrised approach to gravity suggests at least two possibilities for generalisation of extant physics.

Firstly, it provides an interesting way to understand the perennial idea that spacetime is discrete at a fundamental level. Typically, fundamental discreteness is understood as replacing Minkowski spacetime, or perhaps a Riemannian spacetime, with some discretised approximation; such a move inevitably breaks the translational, rotational and boost symmetries of spacetime. The parametrised approach offers an alternative: keep Minkowski spacetime continuous but discretise the base manifold \mc{B}. In this approach, that base manifold is replaced by a finite graph: a collection of points linked by nodes. If the symmetry group is \mc{G} and the typical fibre is \mc{V}, then 
\begin{itemize}
\item Each point $x$ gets its own copy $\mc{V}_x$ of \mc{V};
\item Matter fields assign to each $x$ a point $\varphi(x)$ in $\mc{V}_x$;
\item The gauge connection is an assignment, to each link $x\rightarrow y$, of a structure-preserving map from $V_x$ to $V_y$.
\item A choice of gauge is a map from each $\mc{V}_x$ to (a shared copy of) \mc{V}.
\item Relative to a choice of gauge, matter fields assign a point of \mc{V} to every node; gauge connections assign an element of the automorphism group of \mc{V} to every link.
\end{itemize}
For internal gauge symmetries, this is in fact roughly the formalism already used in lattice gauge theory, though there the lattice is taken to be a lattice of points in Minkowski spacetime. There is also a certain similarity to the lattices used in loop-space quantum gravity, though there the connection takes values in the Lorentz rather than the Poincar\'{e} group.

Secondly, while in parametrised versions of extant field theories the translational part of the Poincar\'{e} group always acts on a pointlike `location field', the general framework requires only a fibre \mc{V} on which that group acts in some way or other. We could, for instance, take \mc{V} to be the space of embeddings of a closed loop into Minkowski spacetime, on which the Poincar\'{e} group acts in the obvious way. A field theory with this fibre would assign a loop to every point on the base space, either lying in the same copy of Minkowski spacetime (for the global theory) or in a local copy. The Poincar\'{e} connection could be defined just as before, giving the relations between loops at infinitesimally separated spacetime points. The connections (if any) between this sort of ``loop-valued field theory'' and the superficially-similar string field theory lie beyond the scope of this paper.

\section{Conclusion}\label{conclusion}

The parametrised approach to field theories allows spacetime and internal symmetries to be treated very much on a par, and in particular allows general relativity and related theories of gravity to be seen as gauge theories of the Poincar\'{e} group in just the way that Yang-Mills theories are gauge theories of internal global symmetries.

This is not to say that there are no disanalogies between the two classes of theory. This paper has ignored entirely the global structure of general relativity, and in particular the various issues surrounding black holes and singularities. However, at the local level it is gratifying to see Poincar\'{e} and Yang-Mills gauge theories as simply applications of the same general formula to different symmetry groups.

This suggests in turn that other differences between the theories --- again, at least at the local level --- can be ascribed to differences in their respective dynamics rather than to differences at the kinematic level. In a companion paper \cite{wallacegaugelagrange} I provide a unified account of the dynamics of the two types of gauge theory from the Lagrangian view and obtain results in support of this suggestion.


\end{document}